\documentclass[12pt]{article}
\usepackage{hyperref}
\usepackage[english]{babel}
\usepackage{amsmath}
\usepackage{latexsym}
\usepackage{amssymb}
\usepackage[all]{xy}
\usepackage[dvips]{graphicx}
\usepackage{epsfig}
\usepackage{psfrag}

\numberwithin{equation}{section} \numberwithin{table}{section}
\numberwithin{figure}{section}

\begin{document}


\begin{titlepage}
  \begin{flushright}
  {\small CQUeST-2009-0306}
  \end{flushright}

  \begin{center}

    \vspace{20mm}

    {\LARGE \bf On Charged
    Lifshitz Black Holes}

    \vspace{10mm}

    Da-Wei Pang$^{\dag}$

    \vspace{5mm}
    {\small \sl $\dag$ Center for Quantum Spacetime, Sogang University}\\
    {\small \sl Seoul 121-742, Korea\\}
    {\small \tt pangdw@sogang.ac.kr}
    \vspace{10mm}

  \end{center}

\begin{abstract}
\baselineskip=18pt We obtain exact solutions of charged
asymptotically Lifshitz black holes in arbitrary $(d+2)$ dimensions,
generalizing the four dimensional solution investigated in
0908.2611[hep-th]. We find that both the conventional Hamiltonian
approach and the recently proposed method for defining mass in
non-relativistic backgrounds do not work for this specific example.
Thus the mass of the black hole can only be determined by the first
law of thermodynamics. We also obtain perturbative solutions in
five-dimensional Gauss-Bonnet gravity. The ratio of shear viscosity
over entropy density and the DC conductivity are calculated in the
presence of Gauss-Bonnet corrections.

\end{abstract}
\setcounter{page}{0}
\end{titlepage}

\pagestyle{plain} \baselineskip=19pt

\tableofcontents

\section{Introduction}
The AdS/CFT correspondence~\cite{Maldacena:1997re, Gubser:1998bc,
Witten:1998qj, Aharony:1999ti} has been extensively studied and
fruitful progress has been made in recent years. In particular, as a
strong-weak duality, it has provided a useful tool for studying
strongly coupled field theories. Moreover, inspired by condensed
matter physics, studies of non-relativistic AdS/CFT correspondence
have been accelerated since last year, which may open a new window
for investigating physical systems in the real world. For reviews
see~\cite{Hartnoll:2009sz}.

In many condensed matter systems near a critical point, there exist
field theories with anisotropic scaling symmetry, that is, temporal
and spatial coordinates scale differently,
\begin{equation}
t~\rightarrow~\lambda^{z}t,~~~~~x^{i}~\rightarrow~\lambda x^{i},
\end{equation}
where $z$ is called the `dynamical exponent'. Until now, there are
two main concrete examples of the gravity duals of non-relativistic
field theories. One is called the Schr\"{o}dinger case, which was
proposed in~\cite{Son:2008ye, Balasubramanian:2008dm} and the finite
temperature generalizations were investigated
in~\cite{Herzog:2008wg, Maldacena:2008wh, Adams:2008wt}. The other
is the Lifshitz case, which was obtained in~\cite{Kachru:2008yh}.
For general $(d+2)-$dimensional spacetime, the dual geometry of
Lifshitz fixed points is given by
\begin{equation}
ds^{2}=L^{2}(-r^{2z}dt^{2}+\frac{dr^{2}}{r^{2}}+r^{2}d\vec{x}^2),
\end{equation}
where $d\vec{x}^2=dx^{2}_{1}+\cdots+dx^{2}_{d}$. The scale
transformation acts as
\begin{equation}
t~\rightarrow~\lambda^{z}t,~~~x~\rightarrow~\lambda
x,~~~r~\rightarrow~\frac{r}{\lambda}.
\end{equation}
When $z=1$, it reduces to the usual AdS metric.

When $d=2$, the above geometry can be obtained from a
$(3+1)-$dimensional gravity coupled with a negative cosmological
constant to abelian two- and three-form field strengths,
\begin{equation}
\label{1eq4} S=\int d^{4}x\sqrt{-g}(R-2\Lambda)-\frac{1}{2}\int\ast
F_{(2)}\wedge F_{(2)}-\frac{1}{2}\int\ast H_{(3)}\wedge
H_{(3)}-c\int B_{(2)}\wedge F_{(2)},
\end{equation}
where $F_{(2)}=dA_{(1)}, H_{(3)}=dB_{(2)}$. To obtain such a
solution the cosmological constant and the gauge field strengths
should take the following values,
\begin{eqnarray}
& &\Lambda=-\frac{z^{2}+z+4}{2L^{2}},~~~c=\frac{\sqrt{2z}}{L},\nonumber\\
& &F_{(2)}=\sqrt{2z(z-1)}Lr^{z-1}dr\wedge
dt,~~~H_{(3)}=2\sqrt{z-1}L^{2}rdr\wedge dx_{1}\wedge dx_{2}.
\end{eqnarray}

However, it is quite difficult to find analytic black hole solutions
which are asymptotically Lifshitz-like. Black hole solutions with
$z=2$ in the above $(3+1)-$dimensional theory were investigated
in~\cite{Danielsson:2009gi} via numerical methods. Lifshitz
topological holes were studied in~\cite{Mann:2009yx} where exact
solutions were found in certain specific examples. Black holes in
asymptotically Lifshitz spacetimes with arbitrary critical exponent
and the corresponding thermodynamic behavior were studied
in~\cite{Bertoldi:2009vn, Bertoldi:2009dt}. Another model of
non-relativistic holography was proposed in~\cite{Taylor:2008tg}
where exact Lifshitz black hole solutions were obtained. For other
recent work on Lifshitz black holes see~\cite{others}. It should be
pointed out that it is very difficult to embed the Lifshitz
background into string theory, although certain string theory duals
of Lifshitz-like fixed points were obtained
in~\cite{Azeyanagi:2009pr}. Recently some no-go theorems for string
duals of non-relativistic Lifshitz-like theories were proposed
in~\cite{Li:2009pf}.

Another type of exact Lifshitz black hole solutions was obtained
in~\cite{Brynjolfsson:2009ct}, where the main purpose was to study
holographic superconductors with Lifshitz scaling. In order to
realize phase transitions, they added a second Maxwell field
$\mathcal{F}_{(2)}$ and a scalar field $\psi$ which was charged
under the new gauge field but neutral under the original gauge
fields $F_{(2)}$ and $H_{(3)}$. It was observed that these two
additional fields were sufficient to ensure a superconducting phase
transition characterized by Lifshitz scaling with $z>1$. Here we
generalize their analysis to arbitrary $(d+2)-$dimensional
spacetime. Rather than studying the holographic superconductor, we
investigate the thermodynamic and hydrodynamic properties of such
black holes. We show that the black hole mass can be obtained by the
first law of thermodynamics, while the other approaches for defining
conserved quantities in non-relativistic backgrounds do not work. We
also obtain the Gauss-Bonnet corrections to five-dimensional charged
Lifshitz black holes by perturbative methods. The ratio of shear
viscosity over entropy density and the DC conductivity are
calculated in the presence of Gauss-Bonnet corrections.

The rest of the paper is organized as follows: We obtain the charged
Lifshitz black hole solutions in general $(d+2)-$ dimensions and
discuss the thermodynamics in Sec. 2. The Gauss-Bonnet corrections
to such black holes in five dimensions are calculated perturbatively
in Sec. 3. We compute the ratio of shear viscosity over entropy
density and the DC conductivity in Sec. 4 and Sec. 5 respectively,
where Gauss-Bonnet corrections are taken into account. We summarize
our main results and discuss related issues in Sec. 6.

\section{The solution and thermodynamics}
In this section we will obtain the exact solutions of charged
Lifshitz black holes and discuss their thermodynamics. First we will
review the charged Lifshitz black hole solution studied
in~\cite{Brynjolfsson:2009ct}. The four-dimensional
action~(\ref{1eq4}) admits the following exact Lifshitz black hole
solutions, whose general form can be summarized as
\begin{equation}
\label{2eq1}
ds^{2}=L^{2}[-r^{2z}f(r)^{2}dt^{2}+\frac{g(r)^{2}}{r^{2}}dr^{2}+r^{2}(d\theta^{2}+\chi(\theta)^2d\phi^2)],
\end{equation}
with
\begin{eqnarray}
& &\chi(\theta)=\sin\theta~~~\rm{if}~~~k=1,\nonumber\\
& &\chi(\theta)=\theta~~~\rm{if}~~~k=0,\nonumber\\
& &\chi(\theta)=\sinh\theta~~~\rm{if}~~~ k=-1,
\end{eqnarray}
where $k=+1, 0, -1$ corresponds to a spherical, flat, and hyperbolic
horizon respectively. A $z=2$ topological black hole with a
hyperbolic horizon was obtained in~\cite{Mann:2009yx}, where
\begin{equation}
f(r)=\frac{1}{g(r)}=\sqrt{1-\frac{1}{2r^{2}}}.
\end{equation}
Another solution is a $z=4$ black hole with a spherical
horizon~\cite{Bertoldi:2009vn}, where
\begin{equation}
f(r)=\frac{1}{g(r)}=\sqrt{1+\frac{1}{10r^{2}}-\frac{3}{400r^{4}}}.
\end{equation}

At first sight, these Lifshitz black holes carry a charge that
couples to the two-form field strength $F_{(2)}$, but actually they
are analogous to Schwarzschild-AdS black holes rather than RN-AdS
black holes. It has been shown in~\cite{Brynjolfsson:2009ct} that
the system cannot undergo superconducting phase transition without
additional field contents. Then a second Maxwell field
$\mathcal{F}_{(2)}$ and a scalar field $\psi$ were incorporated,
which were sufficient to ensure a superconducting phase transition.
It should be pointed out that the scalar $\psi$ was charged under
$\mathcal{F}_{(2)}$ but neutral under $F_{(2)}$ and $H_{(3)}$. The
original Lifshitz gauge fields can be seen as auxiliary
construction, which modify the asymptotic geometry from AdS to
Lifshitz. The authors of~\cite{Brynjolfsson:2009ct} added the
following term to the action
\begin{equation}
S_{\mathcal{F}}=-\frac{1}{2}\int\ast\mathcal{F}_{(2)}\wedge\mathcal{F}_{(2)}.
\end{equation}
The new $z=4$ black hole solution still takes the form
of~(\ref{2eq1}), with
\begin{equation}
f(r)=\frac{1}{g(r)}=\sqrt{1+\frac{k}{10r^{2}}-\frac{3k^{2}}{400r^{4}}-\frac{Q^{2}}{2r^{4}}},
\end{equation}
where the physical charge $Q$ is induced by the second Maxwell field
$\mathcal{F}_{(2)}$.

On the other hand, it has been observed in~\cite{Taylor:2008tg} that
the following action
\begin{equation}
S=\frac{1}{16\pi G_{d+2}}\int
d^{d+2}x\sqrt{-g}(R-2\Lambda-\frac{1}{4}F^{2}-\frac{1}{2}m^{2}A^{2})
\end{equation}
admits $(d+2)-$dimensional Lifshitz spacetime with arbitrary $z$
\begin{equation}
ds^{2}=L^{2}(-r^{2z}dt^{2}+\frac{1}{r^{2}}dr^{2}+r^{2}\sum\limits^{d}_{i=1}dx^{2}_{i})
\end{equation}
as a solution, where the values of the fields are fixed to be
\begin{equation}
\Lambda=-\frac{1}{2L^{2}}[z^{2}+z(d-1)+d^{2}],~~~A_{t}=\sqrt{\frac{2(z-1)}{z}}Lr^{z},~~~
m^{2}=\frac{zd}{L^{2}}.
\end{equation}
However, it is difficult to find Lifshitz black hole solutions in
general dimensions by exploring this action. Then inspired
by~\cite{Brynjolfsson:2009ct}, we add a second Maxwell field
$\mathcal{F}_{(2)}$ to the above effective action in arbitrary
dimensions and we find analytic Lifshitz black hole solutions.

Consider the following action in $(d+2)-$dimensional spacetime
\begin{equation}
S=\frac{1}{16\pi G_{d+2}}\int
d^{d+2}x\sqrt{-g}(R-2\Lambda-\frac{1}{4}F^{2}-\frac{1}{2}m^{2}A^{2}-\frac{1}{4}\mathcal{F}^{2}).
\end{equation}
The corresponding equations of motion are given as follows,
\begin{eqnarray}
\partial_{\mu}(\sqrt{-g}F^{\mu\nu})&=&m^{2}\sqrt{-g}A^{\nu},~~~
\partial_{\mu}(\sqrt{-g}\mathcal{F}^{\mu\nu})=0,\nonumber\\
R_{\mu\nu}&=&\frac{2}{d}\Lambda
g_{\mu\nu}+\frac{1}{2}F_{\mu\lambda}{F_{\nu}}^{\lambda}
+\frac{1}{2}\mathcal{F}_{\mu\lambda}{\mathcal{F}_{\nu}}^{\lambda}+\frac{1}{2}m^{2}A_{\mu}A_{\nu}\nonumber\\
&
&-\frac{1}{4d}F^{2}g_{\mu\nu}-\frac{1}{4d}\mathcal{F}^{2}g_{\mu\nu}.
\end{eqnarray}
Let us take the following ansatz for the black hole metric
\begin{equation}
ds^{2}=L^{2}[-r^{2z}f(r)dt^{2}+\frac{dr^{2}}{r^{2}f(r)}+r^{2}\sum\limits^{d}_{i=1}dx^{2}_{i}].
\end{equation}
It can be seen that the $tt$ and $rr$ components of the Einstein
equations give
\begin{equation}
R^{t}_{t}-R^{r}_{r}=-\frac{(z-1)d}{L^{2}}f(r)=-\frac{m^{2}}{2L^{2}r^{2z}f(r)}A_{t}A_{t}.
\end{equation}

When $z=1$, the above equation leads to the following solution for
the massive vector field,
\begin{equation}
A_{t}=0,~~~F_{rt}=0.
\end{equation}
The second Maxwell field and the cosmological constant are given by
\begin{equation}
\mathcal{F}_{rt}=\frac{qL}{r^{d}},~~~\Lambda=-\frac{d(d+1)}{2L^{2}}.
\end{equation}
The metric of the black hole turns out to be
\begin{equation}
ds^{2}=L^{2}[-r^{2}f(r)dt^{2}+\frac{dr^{2}}{r^{2}f(r)}+r^{2}\sum\limits^{d}_{i=1}dx^{2}_{i}],~~~
f(r)=1-\frac{m}{r^{d+1}}+\frac{q^{2}}{2d(d-1)r^{2d}}.
\end{equation}
This is simply the ordinary planar RN-AdS black hole.

When $z\neq1$, the mass parameter and the cosmological constant
remain the same as the zero-temperature Lifshitz background,
\begin{equation}
m^{2}=\frac{zd}{L^{2}},~~~\Lambda=-\frac{1}{2L^{2}}
[z^{2}+z(d-1)+d^{2}],
\end{equation}
while the massive vector field and the second Maxwell field strength
are given by
\begin{equation}
A_{t}=\sqrt{\frac{2(z-1)}{z}}
Lr^{z}f(r),~~~\mathcal{F}_{rt}=qLr^{z-d-1},
\end{equation}
where
\begin{equation}
f(r)=1-\frac{q^{2}}{2d^{2}r^{z}}.
\end{equation}
It should be emphasized that this charged Lifshitz black hole is
different from the $z=1$ counterparts, as it does not admit extremal
solutions. In this sense it looks more analogous to
Schwarzschild-AdS black holes than RN-AdS black holes. Here the
dynamical exponent $z$ cannot be arbitrary, but has to be fixed by
\begin{equation}
z=2d.
\end{equation}
One can see that in four dimensions the solution agrees with the
$k=0$ black hole solution obtained in~\cite{Brynjolfsson:2009ct},
after redefining the charge parameter $q$. The temperature and
entropy are given by
\begin{equation}
T=\frac{z}{4\pi}r^{z}_{0},~~~S_{\rm
BH}=\frac{L^{d}V_{d}}{4G_{d+2}}r^{d}_{0},
\end{equation}
where $r_{0}^{z}\equiv q^{2}/2d^{2}$ and $V_{d}$ denotes the volume
of the $d-$dimensional spatial part.

Next let us calculate the mass of the Lifshitz black hole. One way
is to consider the Hamiltonian approach, which was proposed
in~\cite{Hawking:1995fd} and was illustrated for four-dimensional
Lifshitz black holes in~\cite{Bertoldi:2009vn, Bertoldi:2009dt}. The
mass is given by the following formula
\begin{equation}
M =-\frac{1}{8\pi G_{d+2}} \int_{S^{\infty}_{t}} d^{d}x N(K-K_0),
\end{equation}
where $N$ is the lapse function, $K$ is the extrinsic curvature of
the Lifshitz black hole metric at constant $r$ and $K_0$ is the
corresponding extrinsic curvature for zero-temperature background
solution with $f(r)=1$. One can obtain
\begin{equation}
M =-\frac{V_{d}L^{d}}{8\pi G_{d+2}} \lim_{r \to \infty } r^{3d}(f-1)
   \sim r^d,
\end{equation}
that is, the mass is divergent. However, one can still define
thermodynamic mass of the solution according to the first law of
black hole thermodynamics,
\begin{equation}
   M = \int TdS_{\rm BH} = \frac{z V_dL^d}{48\pi G_{d+2}} r_0^{3d}.
\end{equation}

Another approach was proposed in~\cite{Ross:2009ar}, where it argued
that the divergences found in~\cite{Bertoldi:2009vn} were due to
using an action which did not include the surface terms necessary to
ensure that the action was finite on shell. However, if we simply
generalize their analysis to general $(d+2)-$dimensional spacetime,
the action can be written as
\begin{eqnarray}
\label{2eq25}
S&=&S_{\rm bulk}+S_{\rm bdy}\nonumber\\
&=&\frac{1}{16\pi G_{d+2}}\int
d^{d+2}x\sqrt{-g}(R-2\Lambda-\frac{1}{4}F_{\mu\nu}F^{\mu\nu}-\frac{1}{2}m^{2}A_{\mu}A^{\mu}
-\frac{1}{4}\mathcal{F}^{2})\nonumber\\
& &+\frac{1}{16\pi G_{d+2}}\int
d^{d+1}\xi\sqrt{-h}(2K-\frac{2d}{L}+f(A_{\alpha}A^{\alpha}))+S_{\rm
deriv}.
\end{eqnarray}
Here $\xi^{\alpha}$ are coordinates on the boundary,
$h_{\alpha\beta}$ is the induced metric and $K_{\alpha\beta}$ is the
extrinsic curvature of the boundary. $S_{\rm deriv}$ is a collection
of terms involving derivatives of the boundary fields. By fixing
$f(A_{\alpha}A^{\alpha})=-(z\alpha/L)\sqrt{-A_{\alpha}A^{\alpha}}$
and taking Dirichlet boundary condition for $\mathcal{A}_{\mu}$, one
can obtain a well-defined variation of the action,
\begin{equation}
\delta S=\int
d^{d+1}\xi(s_{\alpha\beta}h^{\alpha\beta}+s_{\alpha}\delta
A^{\alpha}),
\end{equation}
where
\begin{eqnarray}
\label{2eq27}
s_{\alpha\beta}&=&\sqrt{-h}[(\pi_{\alpha\beta}+\frac{d}{L}h_{\alpha\beta})+\frac{z\alpha}{2L}
(-A_{\alpha}A^{\alpha})^{-1/2}(A_{\alpha}A_{\beta}-A_{\gamma}A^{\gamma}h_{\alpha\beta})]+s^{\rm
deriv}_{\alpha\beta},\nonumber\\
s_{\alpha}&=&-\sqrt{-h}(n^{\mu}F_{\mu\alpha}-\frac{z\alpha}{L}(-A_{\alpha}A^{\alpha})^{-1/2}A_{\alpha})+
s^{\rm deriv}_{\alpha}
\end{eqnarray}
and $\pi_{\alpha\beta}\equiv K_{\alpha\beta}-Kh_{\alpha\beta}$.

The contribution of the second Maxwell field $\mathcal{F}_{\mu\nu}$
to the on-shell action can be estimated as
\begin{equation}
\int\sqrt{-g}\mathcal{F}^{2}\sim\int r^{d-1},
\end{equation}
which is divergent. According to~\cite{Ross:2009ar}, for
asymptotically Lifshitz spacetimes, the energy is given by
\begin{equation}
\mathcal{E}=2{s^{t}}_{t}-s^{t}A_{t}.
\end{equation}
Substituting the black hole solution into~(\ref{2eq27}), the result
turns out to be
\begin{equation}
\mathcal{E}=\lim_{r \to \infty }dLr^{3d}(f(r)-f(r)^{1/2})\sim
r^{d}\rightarrow\infty.
\end{equation}
The divergence implies that the boundary terms in~(\ref{2eq25}) are
a priori insufficient to render the on-shell action finite. We
expect to remove the divergences in further investigations on the
definitions of conserved quantities in asymptotically Lifshitz
spacetimes.

\section{Gauss-Bonnet corrections}
In this section we study Gauss-Bonnet corrections to
five-dimensional charged Lifshitz black holes. Generally speaking,
it is always difficult to obtain exact black hole solutions in
higher derivative gravity, so we will try to find black hole
solutions in Gauss-Bonnet gravity by perturbative methods,
following~\cite{Kats:2007mq}.

First we rewrite the five-dimensional charged Lifshitz black hole
solution as follows,
\begin{eqnarray}
&
&ds^{2}=L^{2}[-r^{2z}f(r)dt^{2}+\frac{dr^{2}}{r^{2}f(r)}+r^{2}(dx^{2}_{1}
+dx^{2}_{2}+dx^{2}_{3})],~~~f(r)=1-\frac{q^{2}}{18r^{z}}\equiv1-\frac{r^{z}_{0}}{r^{z}},\nonumber\\
&
&z=2d=6,~~~m^{2}=\frac{18}{L^{2}},~~~
\Lambda=-\frac{1}{2L^{2}}[z^{2}+(d-1)z+d^{2}]=-\frac{57}{2L^{2}},\nonumber\\
& &A_{t}=\sqrt{\frac{5}{3}}r^{z}f(r),~~~\mathcal{F}_{rt}=qLr^{2}.
\end{eqnarray}
Now consider the following effective action containing Gauss-Bonnet
corrections
\begin{eqnarray}
S&=&\frac{1}{16\pi G_{5}}\int
d^{5}x\sqrt{-g}[R-2\Lambda-\frac{1}{4}F_{\mu\nu}F^{\mu\nu}-\frac{1}{2}m^{2}A_{\mu}A^{\mu}-\frac{1}{4}
\mathcal{F}_{\mu\nu}\mathcal{F}^{\mu\nu}\nonumber\\
& &+\frac{\lambda_{\rm
GB}}{2}L^{2}(R_{\mu\nu\lambda\delta}R^{\mu\nu\lambda\delta}-4R_{\mu\nu}R^{\mu\nu}+R^{2})].
\end{eqnarray}
The equations of motion for the gauge fields remain the same as
before,
\begin{equation}
\label{3eq3}
\partial_{\mu}(\sqrt{-g}F^{\mu\nu})=m^{2}\sqrt{-g}A^{\nu},~~~
\partial_{\mu}(\sqrt{-g}\mathcal{F}^{\mu\nu})=0,
\end{equation}
while the Einstein equations take the following form
\begin{equation}
R_{\mu\nu}-\frac{1}{2}Rg_{\mu\nu}=-\Lambda g_{\mu\nu}+T^{\rm
M}_{\mu\nu}+T^{\rm R}_{\mu\nu},
\end{equation}
with
\begin{eqnarray}
T^{\rm
M}_{\mu\nu}&=&\frac{1}{2}F_{\mu\lambda}{F_{\nu}}^{\lambda}+\frac{1}{2}m^{2}A_{\mu}A_{\nu}+
\frac{1}{2}\mathcal{F}_{\mu\lambda}{\mathcal{F}_{\nu}}^{\lambda}-\frac{1}{8}F^{2}g_{\mu\nu}
-\frac{1}{4}m^{2}A^{2}g_{\mu\nu}-\frac{1}{8}\mathcal{F}^{2}g_{\mu\nu},\nonumber\\
T^{\rm R}_{\mu\nu}&=&\frac{\lambda_{\rm
GB}}{2}L^{2}[\frac{1}{2}g_{\mu\nu}(R_{\mu\nu\lambda\delta}R^{\mu\nu\lambda\delta}
-4R_{\mu\nu}R^{\mu\nu}+R^{2})\nonumber\\&
&-2RR_{\mu\nu}+4R_{\mu\gamma}{R^{\gamma}}_{\nu}+4R^{\gamma\delta}
R_{\gamma\mu\delta\nu}-2R_{\mu\gamma\delta\lambda}{R_{\nu}}^{\gamma\delta\lambda}].
\end{eqnarray}
The Einstein equations can be recasted as
\begin{equation}
\label{3eq6} R_{\mu\nu}=\frac{2}{3}\Lambda g_{\mu\nu}+T^{\rm
M}_{\mu\nu}+T^{\rm R}_{\mu\nu}-\frac{1}{3}(T^{\rm M}+T^{\rm
R})g_{\mu\nu},
\end{equation}
where $T^{\rm M}\equiv g^{\mu\nu}T^{\rm M}_{\mu\nu}$ and $T^{\rm
R}\equiv g^{\mu\nu}T^{\rm R}_{\mu\nu}$.

The ansatz for the metric is given by
\begin{equation}
ds^{2}=\frac{L^{2}}{\rho^{2}}[-e^{2A(\rho)}dt^{2}+e^{-2B(\rho)}d\rho^{2}+dx^{2}_{1}+dx^{2}_{2}+dx^{2}_{3}].
\end{equation}
One can derive the following relations for the components of the
Ricci tensor
\begin{eqnarray}
\label{3eq8} R^{t}_{t}-R^{\rho}_{\rho}&=&\frac{3}{L^{2}}e^{2B(\rho)}
\rho(A^{\prime}(\rho)-B^{\prime}(\rho)),\nonumber\\
\frac{1}{3}(R^{t}_{t}-R^{\rho}_{\rho})-R^{1}_{1}&=&
-\frac{1}{L^{2}}(\frac{e^{2B(\rho)}}{\rho^{4}})^{\prime}\rho^{5},
\end{eqnarray}
where the prime stands for derivative with respect to $\rho$. To
obtain the perturbative solution, we will substitute the field
configurations of the unperturbed solution into $T^{\rm M}_{\mu\nu}$
and $T^{\rm R}_{\mu\nu}$. By combining~(\ref{3eq6})
and~(\ref{3eq8}), we can obtain
\begin{equation}
-\frac{1}{L^{2}}(\frac{e^{2B(\rho)}}{\rho^{4}})^{\prime}\rho^{5}=\frac{2}{3}{T^{\rm
M}}^{t}_{t}+\frac{2}{3}{T^{\rm R}}^{t}_{t}-\frac{2}{3}\Lambda.
\end{equation}
Then the function $e^{2B(\rho)}$ reads
\begin{equation}
e^{2B(\rho)}=f(\rho)(1+\lambda_{\rm
GB}f(\rho)),~~~f(\rho)=1-\frac{q^{2}}{18}\rho^{6}.
\end{equation}
From the first equation of~(\ref{3eq8}), we can obtain
\begin{equation}
A(\rho)=B(\rho)+\frac{L^{2}}{3}\int\frac{1}{\rho}e^{2B(\rho)}(R^{t}_{t}-R^{\rho}_{\rho}).
\end{equation}
After substituting the unperturbed solution, the function
$e^{2A(\rho)}$ is given by
\begin{equation}
e^{2A(\rho)}=e^{2B(\rho)}\rho^{-(z_{0}-1)(2+4\lambda_{\rm
GB})}\exp[\frac{2(z_{0}-1)}{z_{0}}\frac{\lambda_{\rm
GB}q^{2}}{d^{2}}\rho^{z_{0}}],
\end{equation}
where $z_{0}=2d=6$ denotes the dynamical exponent of the unperturbed
solution.

Finally, performing the coordinate transformation $\rho=1/r$, the
perturbed solution in Gauss-Bonnet gravity can be summarized as
follows
\begin{equation}
\label{3eq13}
ds^{2}=L^{2}[-f(r)h(r)r^{2z}dt^{2}+\frac{dr^{2}}{r^{2}f(r)}+r^{2}(dx^{2}_{1}+dx^{2}_{2}+dx^{2}_{3})],
\end{equation}
where
\begin{eqnarray}
f(r)&=&f_{0}(r)(1+\lambda_{\rm
GB}f_{0}(r)),~~~f_{0}(r)=1-\frac{q^{2}}{18r^{6}},\nonumber\\
h(r)&=&\exp[\frac{5\lambda_{\rm GB}q^{2}}{27r^{6}}],
\end{eqnarray}
and
\begin{equation}
z=z_{0}+2\lambda_{\rm GB}(z_{0}-1),~~~z_{0}=2d=6.
\end{equation}
The first-order solutions of the gauge fields can also be obtained
by substituting the perturbed metric into the equations of
motion~(\ref{3eq3}), where we have made the following ansatz
\begin{equation}
A_{t}=\sqrt{\frac{5}{3}}r^{z_{0}}f_{0}(r)+\lambda_{\rm
GB}A_{1t}(r),~~~\mathcal{F}_{rt}=qLr^{2}+\lambda_{\rm
GB}\mathcal{F}_{1rt}.
\end{equation}
The first-order corrections are given by
\begin{eqnarray}
A_{1t}&=&c_{1}(r^{6}-r_{0}^{6})+c_{2}(-\frac{r^{3}}{6r^{6}_{0}}+\frac{{\rm
arctanh}(\frac{r^{3}}{r_{0}^{3}})}{6r^{9}_{0}}(r^{6}-r^{6}_{0}))\nonumber\\
& &+\mathcal{O}(\log(r^{6}-r^{6}_{0})),\nonumber\\
\mathcal{F}_{1rt}&=&\frac{5}{3}qL\frac{r^{6}_{0}}{r^{4}}.
\end{eqnarray}

Here are some remarks on the perturbed black hole solution:
\begin{itemize}
\item The horizon still locates at $r=r_{0}$ for the perturbed black
hole solution.
\item The temperature and entropy are given by
\begin{equation}
T=\frac{1}{4\pi}z_{0}r^{z}_{0}\exp(\frac{5}{3}\lambda_{\rm GB}),~~~
S_{\rm BH}=\frac{1}{4G_{5}}r^{3}_{0}L^{3}V_{3}.
\end{equation}
Note that the Bekenstein-Hawking formula still holds for planar
black holes in Gauss-Bonnet gravity.
\item The $1/N$ effects in non-relativistic gauge-gravity duality
were investigated extensively in~\cite{Adams:2008zk}, where they
argued that the dynamical exponent would be renormalized in higher
derivative gravity except for $z_{0}=1$. Here our results support
their arument.
\end{itemize}
\section{Calculating $\eta/s$}
The AdS/CFT correspondence has provided us a powerful tool for
investigating the hydrodynamic properties of strongly coupled field
theories. One remarkable progress is the calculation of shear
viscosity in the dual gravity side. It has been found that the ratio
of shear viscosity over entropy density is $1/4\pi$ for a large
class of CFTs with Einstein gravity duals in the large N limit.
Therefore, it was conjectured that $1/4\pi$ is a universal lower
bound for all materials, which is the so-called
Kovtun-Son-Starinets(KSS) bound~\cite{Kovtun:2004de}. However,
in~\cite{Kats:2007mq, Brigante:2007nu, Brigante:2008gz} it was
observed that in $R^{2}$ gravity such a lower bound was violated and
a new lower bound $4/25\pi$ was proposed by considering the
causality of the dual field theory. Corrections to $\eta/s$ for
various examples were nicely investigated in~\cite{Myers:2008yi,
Buchel:2008vz}. For more developments, see e.g.~\cite{more}.

It was conjectured in~\cite{Brustein:2007jj} that the shear
viscosity is fully determined by the effective coupling of the
transverse gravitons on the horizon in the dual gravity description.
This was confirmed in~\cite{Iqbal:2008by} via the scalar membrane
paradigm and in~\cite{Cai:2008ph} by calculating the on-shell action
of the transverse gravitons. However, the latter formalism is not
covariant under coordinate transformations, then the choice of
coordinate system of the background black hole geometry affects the
form of the action of the transverse gravitons. A new formalism was
proposed in~\cite{Cai:2009zv}, where a three-dimensional effective
metric $\tilde{g}_{\mu\nu}$ was introduced and the transverse
gravitons were minimally coupled to this new effective metric. The
action in this new formalism can take a covariant form. Similar
discussions on this issue were also presented
in~\cite{Banerjee:2009wg}.

The shear viscosity is given by the Kubo formula
\begin{equation}
\eta=\lim\limits_{\omega\rightarrow0}\frac{1}{2\omega
i}(G^{A}_{x_{1}x_{2},x_{1}x_{2}}(\omega,0)-G^{R}
_{x_{1}x_{2},x_{1}x_{2}}(\omega,0)),
\end{equation}
where the retarded Green's function $G^{R}_{\mu\nu,\lambda\rho}$ is
defined by
\begin{equation}
G^{R}_{\mu\nu,\lambda\rho}=-i\int d^{4}xe^{-ik\cdot x}\theta(t)
<[T_{\mu\nu}(x),T_{\lambda\rho}(0)]>,
\end{equation}
and the advanced Green's function satisfies
$G^{A}_{\mu\nu,\lambda\rho}(k)={G^{R}_{\mu\nu,\lambda\rho}(k)}^{\ast}$.
According to the field-operator correspondence, such Green's
functions can be calculated through the effective action of the
gravitons of the dual gravity theory.

Consider tensor perturbation $h_{12}=h_{12}(t,u,z)$, where $u$ is
the radial coordinate and the momentum of the perturbation points
along the $x_{3}\equiv z$ axis. Then we denote $\phi=h^{1}_{2}$ and
write $\phi$ as $\phi(t,u,z)=\phi(u)e^{-i\omega t+ipz}$. If the
transverse gravitons can be decoupled from other perturbations, the
effective bulk action of the transverse gravitons can be written in
a general form
\begin{equation}
\label{4eq3} S=\frac{V_{1,2}}{16\pi G}(-\frac{1}{2})\int
d^{3}x\sqrt{-\tilde{g}}(\tilde{K}(u)\tilde{g}^{MN}
\tilde{\nabla}_{M}\phi\tilde{\nabla}_{N}\phi+m^{2}\phi^{2})
\end{equation}
up to some total derivatives. Here $\tilde{g}_{MN}, {M,N=t,u,z}$ is
a three-dimensional effective metric, $m$ is an effective mass and
$\tilde{\nabla}_{M}$ is the covariant derivative using
$\tilde{g}_{MN}$. Notice that $\phi$ is a scalar in the three
dimensions $t,u,z$, while it is not a scalar in the whole five
dimensions. As we have assumed that $\phi=h^{1}_{2}(t,u,z)$, the
effective action for $\phi$ can be seen as a deduced
three-dimensional action where the other two irrelevant directions
can be integrated out. In this sense it is not the usual dimensional
reduction. The three-dimensional effective action itself is general
covariant and $\tilde{K}(u)$ is a scalar under general coordinate
transformations. In the following we will use $g_{\mu\nu}$ to denote
the whole five-dimensional background.

By performing the coordinate transformation $u=r_{0}^{3}/r^{3}$, the
perturbative black hole solution in Gauss-Bonnet
gravity~(\ref{3eq13}) can be rewritten as
\begin{equation}
\label{4eq4}
ds^{2}=L^{2}[-(1-u)F(u)dt^{2}+\frac{du^{2}}{(1-u)G(u)}+\frac{r^{2}_{0}}{u^{2/3}}
(dx^{2}_{1}+dx^{2}_{2}+dx^{2}_{3})],
\end{equation}
where
\begin{eqnarray}
& &F(u)=(1+u)(1+\lambda_{\rm
GB}(1-u^{2}))h(u)r^{2z}_{0}u^{-2z/3},\nonumber\\
& &h(u)=\exp[\frac{5}{27}\lambda_{\rm
GB}u^{2}],~~~G(u)=9(1+u)(1+\lambda_{\rm
GB}(1-u^{2}))u^{2},\nonumber\\
& &z=z_{0}+2\lambda_{\rm GB}(z_{0}-1),~~~z_{0}=2d=6.
\end{eqnarray}
Notice that here the horizon locates at $u=1$.
Following~\cite{Cai:2009zv}, we write down the action of the
transverse gravitons in momentum space
\begin{eqnarray}
S&=&\frac{V_{1,2}}{16\pi
G}(-\frac{1}{2})\int\frac{dwdp}{(2\pi)^{2}}du\sqrt{-\tilde{g}}[
\tilde{K}(u)(\tilde{g}^{uu}\phi^{\prime}\phi^{\prime}\nonumber\\
&
&+w^{2}\tilde{g}^{tt}\phi^{2}+p^{2}\tilde{g}^{zz}\phi^{2})+m^{2}\phi^{2}],
\end{eqnarray}
where
\begin{eqnarray}
&
&\phi(t,u,z)=\int\frac{dwdp}{(2\pi)^{2}}\phi(u;k)e^{-iwt+ipz},\nonumber\\
& &k=(w,0,0,p),~~~\phi(u;-k)=\phi^{\ast}(u;k),
\end{eqnarray}
and the prime denotes derivative with respect to $u$. The subsequent
steps are similar to those exhibited in~\cite{Cai:2009zv}, which
will not be shown explicitly. We can find that here we still have
the following formula for $\eta$
\begin{equation}
\label{4eq8} \eta=\frac{1}{16\pi
G}(\sqrt{\tilde{g_{zz}}}\tilde{K}(u))|_{u=1}.
\end{equation}

Next we calculate the effective action of the transverse gravitons
in the background~(\ref{4eq4}). By checking directly from the
first-order Einstein equations it can be found that the transverse
gravitons can get decoupled from other perturbations. Then the
effective action of the transverse gravitons can be obtained by
keeping the quadratic terms of $\phi$ in the original action and it
can be written in the form of~(\ref{4eq3}) with the effective
three-dimensional metric
\begin{equation}
\tilde{g}^{uu}=(1+\frac{\lambda_{\rm
GB}}{2}\frac{Ag^{\prime}_{tt}g^{uu}}{ug_{tt}})g^{uu},
\end{equation}
\begin{equation}
\tilde{g}^{tt}=[1+\frac{\lambda_{\rm GB}}{2}(\frac{Ag^{\prime
uu}}{u}-\frac{(A^{2}+2A)g^{uu}}{u^{2}})]g^{tt},
\end{equation}
\begin{equation}
\label{4eq11} \tilde{g}^{zz}=[1+\frac{\lambda_{\rm
GB}}{2}(\frac{g^{\prime2}_{tt}g^{uu}}{g^{2}_{tt}}
-\frac{g^{\prime}_{tt}g^{\prime
uu}}{g_{tt}}-\frac{2g^{uu}g^{\prime\prime}_{tt}}{g_{tt}})]g^{zz},
\end{equation}
where $A\equiv 4/z_{0}=2/3$.

In fact, the effective action of the transverse gravitons can also
be written as
\begin{equation}
S=\frac{1}{16\pi G}(-\frac{1}{2})\int
d^{5}x\sqrt{-g}\tilde{g}^{\mu\nu}\partial_{\mu}\phi\partial_{\nu}\phi,
\end{equation}
where $\tilde{g}^{\mu\nu}=\tilde{g}^{\mu\nu}$ for $\mu,\nu=t,u,z$
and $\tilde{g}^{\mu\nu}=g^{\mu\nu}$ for $\mu,\nu=x_{1},x_{2}$. Then
$\tilde{K}(u)=\sqrt{-g}/\sqrt{-\tilde{g}}$. Finally, by
substituting~(\ref{4eq11}) into~(\ref{4eq8}) and recalling the fact
that the Bekenstein-Hawking area law still holds for planar black
holes in Gauss-Bonnet gravity, we have
\begin{eqnarray}
\frac{\eta}{s}&=&\frac{1}{4\pi}[1-\frac{A}{2}\lambda_{\rm
GB}G(1)]\nonumber\\
&=&\frac{1}{4\pi}(1-6\lambda_{\rm GB}).
\end{eqnarray}
Here are some remarks on this result:
\begin{itemize}
\item The Gauss-Bonnet corrections to asymptotically Lifshitz black
holes and the higher order corrections to $\eta/s$ were calculated
in~\cite{Pang:2009ky}, where the result turned out to be
\begin{equation}
\frac{\eta}{s}=\frac{1}{4\pi}[1-(z_{0}+3)\lambda_{\rm GB}].
\end{equation}
It can be seen that the above expression reduces to the result
obtained in~\cite{Brigante:2007nu} when $z_{0}=1$. However, here we
cannot reproduce the same result as the dynamical exponent $z_{0}$
is not a free parameter but is fixed by the number of spatial
coordinates.
\item
$\lambda_{\rm GB}$ should have an upper bound $1/6$ to ensure a
non-vanishing $\eta/s$. The upper bound of $\lambda_{\rm GB}$ was
discussed in~\cite{Ge:2009eh} where it was found to be $1/4$ by the
constraints of causality and stability. Here a similar upper bound
in non-relativistic theory requires further understanding. As the
dual field theory is non-relativistic, causality cannot be served as
a constraint but stability may still work.
\item In the literatures discussing the ratio of shear viscosity
over entropy density in higher derivative theory of gravity, the new
lower bound of $\eta/s$--$4/25\pi$--can be obtained by considering
the causality of the boundary field theory. However, here we cannot
take such a constraint as the dual field theory is non-relativistic,
which is similar to the case discussed in~\cite{Pang:2009ky}.
However, we expect that the issue of stability may still exert some
constraint on $\eta/s$.
\end{itemize}

\section{Conductivity}
In this section we calculate the DC conductivity of the Lifshitz
black holes. Firstly, in order to obtain a well-defined result of
the `electrical' conductivity, we imagine coupling the CFT current
to an external or auxiliary vector field, according
to~\cite{CaronHuot:2006te, Kovtun:2008kx}. The global $U(1)$
symmetry of the theory is gauged with a small coupling $e$. To
leading order in $e$, the effects of the auxiliary gauge field can
be neglected and the conductivity can be determined from the
original CFT. Then the DC conductivity is given by
\begin{equation}
\label{5eq1}
\sigma=-\lim_{\omega\rightarrow0}\frac{e^{2}}{\omega}{\rm
Im}G^{R}_{x,x}(\omega, {\bf k}=0),
\end{equation}
where the retarded correlation function of the global $U(1)$
symmetry currents is given by a Kubo formula similar to the one for
shear viscosity,
\begin{equation}
\label{5eq2} G^{R}_{x,x}(\omega, {\bf k}=0)=-i\int dtd{\bf
x}e^{i\omega t}\theta(t)<[J_{x}(x), J_{x}(0)]>.
\end{equation}

Due to the existence of the non-vanishing component
$\mathcal{A}_{t}$ of the bulk gauge field in the background, the
perturbation of the gauge field $\mathcal{A}_{x}$ gets coupled to
the shear mode graviton, i.e., metric perturbations of the form
$h_{xi}$. However, the contribution of $h_{xi}$ can be integrated
out by imposing gauge invariance upon the two sets of perturbations.
Then we can obtain an action that only contains the
$\mathcal{A}_{x}$ fluctuation. Therefore, the most convenient way to
perform the calculations is the effective action approach. Such an
approach was applied in~\cite{Myers:2009ij}, where the conductivity
for five-dimensional charged planar AdS black holes was computed in
the presence of a general set of four-derivative interactions. We
will follow their approach to calculate the conductivity of the
charged Lifshitz black holes.

The perturbations of the metric and the gauge field are given by
\begin{eqnarray}
\label{5eq3}
{h_{t}}^{x}&=&\int\frac{d^{4}k}{(2\pi)^{4}}t_{k}(u)e^{-i\omega
t+ikz},\nonumber\\
{h_{u}}^{x}&=&\int\frac{d^{4}k}{(2\pi)^{4}}h_{k}(u)e^{-i\omega
t+ikz},\nonumber\\
\mathcal{A}_{x}&=&\int\frac{d^{4}k}{(2\pi)^{4}}a_{k}(u)e^{-i\omega
t+ikz},
\end{eqnarray}
where we have denoted $x_{1}\equiv x$ and $x_{3}\equiv z$. Let us
consider the leading order solution first. After performing
coordinate transformation $u=r^{3}_{0}/r^{3}$, the metric turns out
to be
\begin{equation}
ds^{2}=L^{2}[-\frac{r^{12}_{0}}{u^{4}}f_{0}(u)dt^{2}+\frac{du^{2}}{9u^{2}f_{0}(u)}+
\frac{r^{2}_{0}}{u^{2/3}}(dx^{2}_{1}+dx^{2}_{2}+dx^{2}_{3})],
\end{equation}
where $f_{0}(u)=1-u^{2}$. The horizon locates at $u=u_{0}=1$. As a
gauge choice, we would like to set the perturbation ${h_{u}}^{x}$ to
be zero. By comparing the $xx$ and $ux$ components of the Einstein
equations, we can arrive at the following constraint
\begin{equation}
g_{xx}t^{\prime}_{k}=-\mathcal{A}^{\prime}_{t}a_{k},
\end{equation}
which is the same as that in~\cite{Myers:2009ij}.

After taking the above constraint and the gauge choice $h_{k}(u)=0$,
the quadratic action for $a_{k}$ takes the following form
\begin{equation}
\label{5eq6} \tilde{I}^{(2)}_{a}=\frac{1}{16\pi
G_{5}}\int\frac{d^{4}k}{(2\pi)^{4}}du(N(u)a^{\prime}_{k}a^{\prime}_{-k}
+M(u)a_{k}a_{-k}),
\end{equation}
where
\begin{equation}
N(u)=-3Lr^{7}_{0}u^{-4/3}f_{0}(u),~~~M(u)=\frac{L\omega^{2}u^{2/3}}{3r^{5}_{0}f_{0}(u)}-
\frac{3u^{8/3}}{Lr^{5}_{0}}\mathcal{A}^{\prime2}_{t}.
\end{equation}
Such an effective action can either be obtained by expanding the
original action up to quadratic order of $a_{k}$, or by directly
considering the Maxwell equation of the perturbations. The equation
of motion for $a_{k}$ can be re-expressed as
\begin{equation}
\partial_{u}j_{k}(u)=\frac{1}{8\pi G_{5}}M(u)a_{k}(u),
\end{equation}
where
\begin{equation}
j_{k}(u)\equiv\frac{\delta\tilde{I}^{(2)}_{a}}{\delta
a^{\prime}_{-k}}=\frac{1}{8\pi G_{5}}N(u)a^{\prime}_{k}(u).
\end{equation}
In the near horizon region, the equation of motion can be solved by
taking the ansatz
$$a_{k}(u)=Cf_{0}(u)^{\beta}.$$
It turns out that by requiring regularity at the horizon, we still
have $\beta=\pm i\omega/4\pi T$ as usual. The infalling boundary
condition fixes $\beta=-i\omega/4\pi T$.

Next, rather than solving the equation of motion for $a_{k}$ as in
the conventional cases, we follow the prescription described
in~\cite{Myers:2009ij}. According to~\cite{Iqbal:2008by}, the
condition of regularity at the horizon $u=u_{0}$ requires
\begin{equation}
\label{5eq10} j_{k}(u_{0})=-i\omega\lim_{u\rightarrow
u_{0}}\frac{N(u)}{8\pi
G_{5}}\sqrt{-\frac{g_{uu}}{g_{tt}}}a_{k}(u_{0})+\mathcal{O}(\omega^{2}),
\end{equation}
where we are expanding in small $\omega$ with the zero-frequency
limit of~(\ref{5eq1}) in mind. The flux factor can be identified by
evaluating the on-shell action, which simply yields
\begin{equation}
2\mathcal{F}_{k}=j_{k}(u)a_{-k}(u)
\end{equation}
Then one can obtain the Green's function~(\ref{5eq2}) by evaluating
the flux factor at the asymptotic boundary. The DC conductivity is
given by a formula which is analogous to the one for shear viscosity
\begin{equation}
\label{5eq12}
\sigma=\lim_{u,\omega\rightarrow0}\frac{e^{2}}{\omega}{\rm
Im}[\frac{2\mathcal{F}_{k}}{a_{k}(u)a_{-k}(u)}]_{{\bf
k}=0}=e^{2}\lim_{u,\omega\rightarrow0}\frac{{\rm
Im}[j_{k}(u)a_{-k}(u)]}{\omega a_{k}(u)a_{-k}(u)}|_{{\bf k}=0}.
\end{equation}

In the computations of shear viscosity via the membrane
paradigm~\cite{Iqbal:2008by}, the evolution of the canonical
momentum in the radial direction is completely trivial in the
low-frequency limit. Then it can be evaluated at any radial position
and it is calculated at the horizon. However, here neither $j_{k}$
nor $\omega a_{k}$ has a trivial evolution along the radial
direction even in the low-frequency limit. It can be seen from the
equation of motion that the effective mass $M(u)$ no longer vanishes
in the low-frequency limit due to the existence of
$\mathcal{A}^{\prime}_{t}$, so the equation of motion still produces
a nontrivial flow along the radial direction. However, it was
observed in~\cite{Myers:2009ij} that the radial evolution of the
numerator of~(\ref{5eq12}) is trivial,
\begin{equation}
\partial_{u}{\rm Im}[j_{k}(u)a_{-k}(u)]={\rm Im}[f_{1}(u)a_{k}a_{-k}
+f_{2}(u)j_{k}j_{-k}]=0,
\end{equation}
which is independent of taking the low-frequency limit. Therefore we
can evaluate the numerator at the horizon. Moreover, $j_{k}(u)$
should satisfy the regularity condition given in~(\ref{5eq10}) at
the horizon. The conductivity can be expressed as
\begin{equation}
\sigma=\frac{e^{2}}{8\pi
G_{5}}\kappa^{A}_{2}(u_{0})\frac{\mathcal{N}(u_{0})}{\mathcal{N}(0)}|_{{\bf
k}=0},
\end{equation}
where
\begin{equation}
\kappa^{A}_{2}(u)=-N(u)\sqrt{-\frac{g_{uu}}{g_{tt}}}=Lr_{0}u^{-1/3},~~~
\mathcal{N}(u)=a_{k}(u)a_{-k}(u).
\end{equation}

As emphasized in~\cite{Myers:2009ij}, $\mathcal{N}(u)$ is real and
so independent of $\omega$ up to $\mathcal{O}(\omega^{2})$. It also
means that at this order $\mathcal{N}(u)$ is completely regular at
the horizon. Then we can solve $a_{k}(u)$ by imposing regularity at
the horizon and setting $\omega$ to zero. Thus the equation of
motion can be simplified
\begin{equation}
\partial_{u}(u^{-4/3}f_{0}(u)a^{\prime}_{k}(u))=2u^{-4/3}a_{k}(u),
\end{equation}
whose solution is given by
\begin{equation}
a_{k}(u)=a_{k}(0)_{2}F_{1}[-\frac{1+i\sqrt{71}}{12},
-\frac{1-i\sqrt{71}}{12}, -\frac{1}{6}, u^{2}].
\end{equation}
The approximate solution turns out to be
\begin{equation}
a_{k}(u)=a_{0}(1-3u^{2}).
\end{equation}
Finally, in the leading order background, the conductivity is
\begin{equation}
\sigma=\frac{e^{2}}{2\pi G_{5}}Lr_{0}.
\end{equation}
Recall that the temperature is given by $T=z_{0}r^{z_{0}}/4\pi$, the
expression for the conductivity can be rewritten as
\begin{equation}
\sigma=\frac{e^{2}}{2\pi
G_{5}}(\frac{\pi}{z_{0}})^{\frac{1}{z_{0}}}T^{\frac{1}{z_{0}}}.
\end{equation}
We can see that when $z_{0}=1$(although $z_{0}=6$ is fixed here),
the above result reproduces the conventional relation $\sigma\sim
T$. However, the physical meaning of the non-trivial dependence on
$z_{0}$ in the relation between $\sigma$ and $T$ requires further
interpretation.

The calculations in the first-order background are straightforward.
The perturbative black hole solution in Gauss-Bonnet gravity reads
\begin{eqnarray*}
&
&ds^{2}=L^{2}[-(1-u)F(u)dt^{2}+\frac{du^{2}}{(1-u)G(u)}+\frac{r^{2}_{0}}{u^{2/3}}
(dx^{2}_{1}+dx^{2}_{2}+dx^{2}_{3})],\nonumber\\
& &F(u)=(1+u)(1+\lambda_{\rm
GB}(1-u^{2}))h(u)r^{2z}_{0}u^{-2z/3},\nonumber\\
& &h(u)=\exp[\frac{5}{27}\lambda_{\rm
GB}u^{2}],~~~G(u)=9(1+u)(1+\lambda_{\rm
GB}(1-u^{2}))u^{2},\nonumber\\
& &z=z_{0}+2\lambda_{\rm GB}(z_{0}-1),~~~z_{0}=2d=6
\end{eqnarray*}
and the perturbations are given by~(\ref{5eq3}). We still expand the
action up to quadratic terms for $a_{k}, t_{k}, h_{k}$ and obtain a
constraint by setting $h_{k}=0$. Plugging this constraint back into
the action and keeping terms linear in $\lambda_{\rm GB}$ we can
obtain an action of the form~(\ref{5eq6}). Here the equation of
motion for $a_{k}$ becomes much more complicated and it is quite
difficult to obtain an analytic result. Then we solve the equation
of motion up to leading order of $\lambda_{\rm GB}$. The final
result is given by
\begin{equation}
\sigma=\frac{e^{2}}{2\pi G_{5}}Lr_{0}(1+\frac{5}{3}\lambda_{\rm
GB}).
\end{equation}
\section{Summary and discussion}
We study exact solutions of charged Lifshitz black holes and the
corresponding thermodynamic and hydrodynamic properties in this
paper. We generalize the four-dimensional solution obtained
in~\cite{Brynjolfsson:2009ct} to arbitrary $(d+2)-$dimensional cases
by adding a second Maxwell field in the effective action. The black
hole solutions we find exhibit an unusual thermodynamic behavior. We
cannot obtain a finite mass via the Hamiltonian
approach~\cite{Hawking:1995fd} or the recently proposed
prescriptions for non-relativistic
backgrounds~\cite{Bertoldi:2009vn, Bertoldi:2009dt, Ross:2009ar}.
However, the mass of the black hole can still be obtained by the
first law of thermodynamics. We also obtain the five-dimensional
black hole solutions in Gauss-Bonnet gravity by perturbative
methods. Furthermore, we calculate the ratio of shear viscosity over
entropy density and the DC conductivity in the presence of
Gauss-Bonnet corrections.

Here the ratio of shear viscosity over entropy density in
Gauss-Bonnet gravity also violates the conjectured KSS bound, which
is similar to the known examples. However, unlike the case studied
in~\cite{Pang:2009ky}, here we cannot reproduce the result for
planar AdS black holes in Gauss-Bonnet gravity, as the dynamical
exponent $z_{0}$ is not arbitrary but fixed by the number of the
spatial coordinates. In relativistic cases, a new lower bound can be
obtained by taking the causality of the dual boundary field theory
as a constraint. However, for the non-relativistic cases, the
causality of the boundary field theory cannot be taken as a
constraint, as the speed of light tends to infinity. We expect that
some other criterions which are valid both in relativistic and
non-relativistic backgrounds, such as stability, unitarity and
locality, may introduce a new lower bound for non-relativistic
cases.

When we rewrite the conductivity as a function of the temperature
$T$, it shows that the conductivity is proportional to a non-trivial
power of $T$, where the power is determined by $z_{0}$. The
dependence on $z_{0}$ requires further understanding, although it
reduces to the conventional case $\sigma\sim T$ when $z_{0}=1$.

The definition of the conductivity takes a slightly different form
in~\cite{Myers:2009ij},
\begin{equation}
\sigma=-\lim_{\omega\rightarrow0}\frac{e^{2}L^{2}_{\ast}}{\omega}{\rm
Im}G^{R}_{x,x}(\omega, {\bf k}=0),
\end{equation}
where $L_{\ast}$ is some scale to make sure that the chemical
potential has the appropriate unit of energy after rescaling
$\mathcal{A}_{\mu}=L_{\ast}\tilde{\mathcal{A}}_{\mu}$. The ratio
$\sigma T^{2}/\eta e^{2}$ was also examined in~\cite{Myers:2009ij}.
If we take this definition and fix $L_{\ast}=\pi L$ as they did, the
ratio is given by
\begin{equation}
\frac{\sigma T^{2}}{\eta e^{2}}=\frac{1}{2}z^{2}_{0}r^{2z_{0}-2}.
\end{equation}
It still has a non-trivial dependence on $z_{0}$. When $z_{0}=1$,
this ratio is $1/2$, which seems to satisfy the upper bound $\sigma
T^{2}/\eta e^{2}=1$. However, as pointed out
in~\cite{Kovtun:2008kx}, the definition of $\sigma$ (or $\eta$)
involves an arbitrary choice of normalization for the corresponding
current. So in order to find a universal bound on conductivity, it
is more natural to incorporate a quantity which is independent of
the normalization.

Recently there have been several interesting papers on the transport
coefficients for extremal black holes in higher derivative gravity,
e.g.~\cite{Edalati:2009bi, Paulos:2009yk, Cai:2009zn,
Chakrabarti:2009ht, Banerjee:2009ju}. One conclusion is that for
extremal black holes the DC conductivity is always zero, even in
higher derivative gravity theories. The near horizon geometry of
extremal black holes contains an $AdS_{2}$ part, which plays an
important role in the calculations. Here the DC conductivity is also
zero if we take $T=0$. However, the behavior of charged Lifshitz
black holes is more analogous to Schwarzschild-AdS black holes
rather than RN-AdS black holes thus we cannot have $AdS_{2}$ near
horizon geometry. We believe that for ``real'' extremal charged
Lifshitz black holes, the conclusion will be the same as the
relativistic counterparts, as they will exhibit similar near horizon
structure.

\bigskip \goodbreak \centerline{\bf Acknowledgements}
\noindent We thank Rong-Gen Cai for collaboration at the initial
stage of this work and for valuable comments on the manuscript. Part
of this work was done during DWP's visit to Department of Physics,
National Central University and we would like to thank Chiang-Mei
Chen for warm hospitality and helpful discussion. This work was
supported by the Korea Science and Engineering Foundation(KOSEF)
grant funded by the Korea government(MEST) through the Center for
Quantum Spacetime(CQUeST) of Sogang University with grant number
R11-2005-021.



\end{document}